\documentclass[fleqn,twoside]{article}
\usepackage{espcrc2}
\usepackage{latexsym}
\usepackage{epsfig}

\usepackage{graphicx}

\newcommand{\be}{\begin{equation}}
\newcommand{\ee}{\end{equation}}
\newcommand{\ba}{\begin{eqnarray}}
\newcommand{\ea}{\end{eqnarray}}

\def\lsi{\raise0.3ex\hbox{$<$\kern-0.75em\raise-1.1ex\hbox{$\sim$}}}
\def\gsi{\raise0.3ex\hbox{$>$\kern-0.75em\raise-1.1ex\hbox{$\sim$}}}

\newcommand{\fzs}{{\rm FZS}}
\newcommand{\xy}{{\rm XY}}
\newcommand{\half}{\displaystyle \frac{1}{2}}

\title{The inverted XY universality of the superconductivity phase 
	transition}

\author{T. Neuhaus\address{Finkenweg 15, D-33824 Werther, Germany},
	A. Rajantie\address{DAMTP, CMS, University of Cambridge,
	Cambridge CB3 0WA, United Kingdom} and
	K. Rummukainen\address{NORDITA, Blegdamsvej 17, DK-2100
	Copenhagen \O, Denmark {\em and} Department of Physics,
	P.O.Box 64, 00014 University of Helsinki, Finland}\\
\vspace*{-2cm}
\mbox{}\hfill NORDITA-2002-60 HE, HIP 2002-39/TH \\
\vspace{1.7cm}
	}

\begin{document}
\begin{abstract}
It has been conjectured that the phase transition in the
Ginzburg-Landau theory is dual to the XY model transition.  We study
numerically a particular limit of the GL theory where this duality
becomes exact, clarifying some of the problems encountered in standard
GL theory simulations.  This may also explain the failure of the
superconductor experiments to observe the XY model scaling.
\end{abstract}

\maketitle

\section{INTRODUCTION}

The (3-d) U(1) gauge + Higgs (Ginzburg-Landau, GL) theory is an
effective theory for the superconductor-insulator phase transition.
Despite the formal simplicity of the GL theory, and numerous
analytical and numerical studies, the universal properties of the
transition have not yet been fully resolved.  Theoretical {\em
duality arguments} \cite{duality} suggest that the phase transition is
in the 3d XY model universality class, but with an inverted
temperature axis.  Thus, the symmetric and broken phases in the XY
model correspond to the insulator and superconducting phases of the
superconductor.

The duality gives concrete predictions for the behaviour of several
critical observables in superconductors.  For example, the
{\em{Abrikosov vortex tension ${\cal T}$}} should scale with the XY
model exponent $\nu_\xy = 0.6723$\footnote{Accurate value of
$\nu_\xy$ can be found in Ref.~\cite{Hasenbusch}}.  Experimentally, it is easier to
access the {{\em penetration depth $\lambda$}} (or the inverse photon
mass), which is also argued to scale with the XY exponent $\nu'=\nu_\xy$.

However, the XY universality has {\em not} been unambiguously
observed.  Two different high-$T_c$ YBa$_2$Cu$_3$O$_{7-\delta}$
experiments \cite{experiment1} report $\nu'$ in the range
$\approx 0.3 \ldots 0.5$.  Monte Carlo simulations of the
GL model favour $\nu' \sim 0.3$ \cite{us}.\footnote{ In
the London limit (fixed length Higgs), where the duality is on a
firmer footing, simulations appear to give $\nu'\approx 0.67$
\cite{Londonlimit}.}

One reason for the confusing results both in experiments and numerical
simulations is that the duality is expected to be valid only in a very
narrow temperature range around the critical temperature, requiring
extreme precision and large volumes in the measurements.  Moreover,
the duality relates ``simple'' observables (like the field expectation
value in the XY model) to non-local observables in the dual theory
(vortex network in the GL theory).  This makes it very difficult to
know whether the problems are caused by insufficient resolution
near the critical temperature, or by the difficult nature of the
observables themselves.

In order to gain insight into this problem we study a special limit of
the GL theory, the {\em{frozen superconductor}} (FZS) (an
integer-valued gauge theory), which is {\em exactly} dual to the
XY-Villain model at all temperatures.  Thus, the transition in FZS is
bound to be in the XY model universality class.  Studying the critical
quantities of FZS can shed light on the problems faced in both the
GL theory simulations and superconductor experiments.
Detailed results are published in \cite{results}.

\begin{figure*}
\vspace{-2mm}
\centerline{\hfill\psfig{file=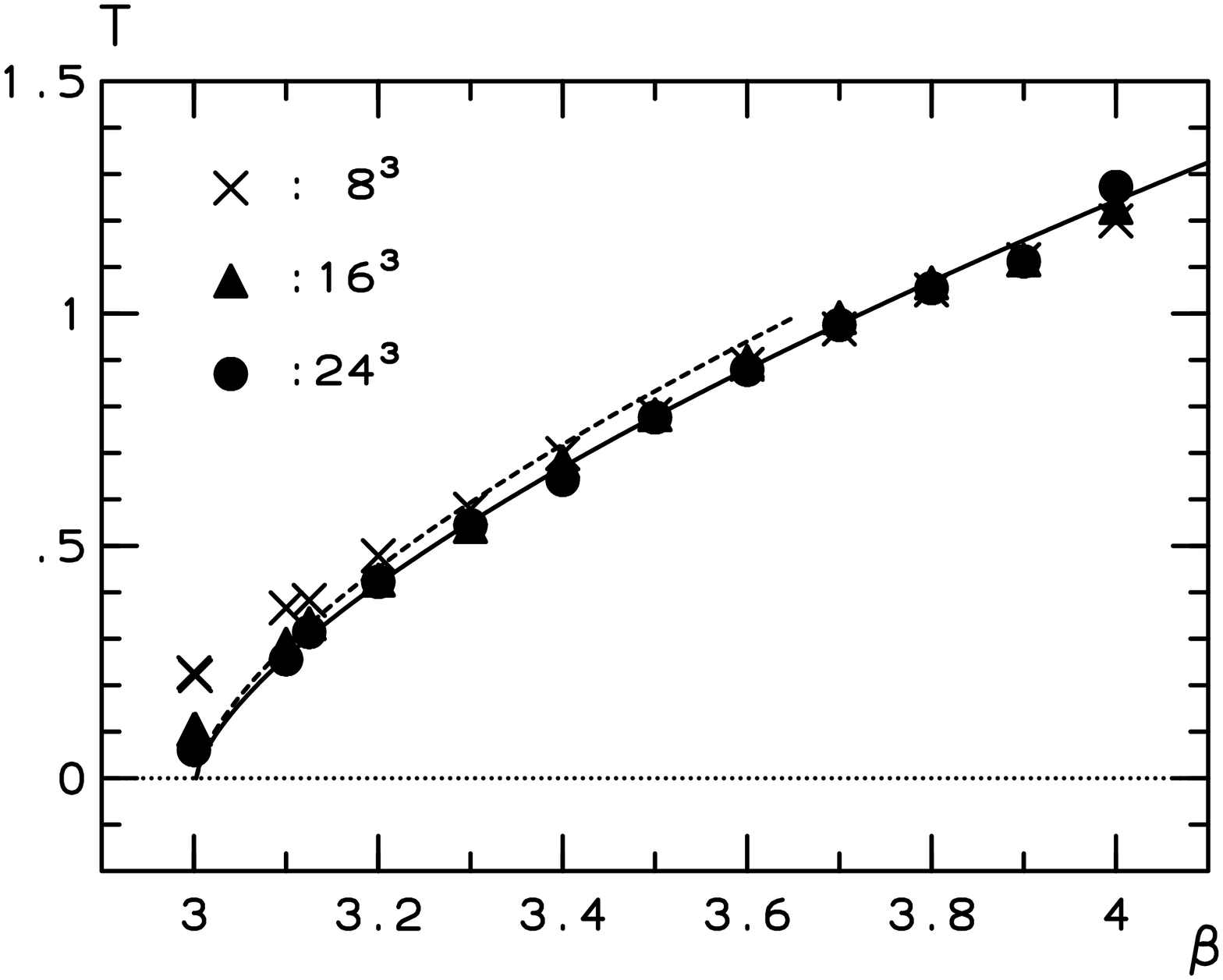,angle=270,width=6.7cm}
\hfill\hfill\psfig{file=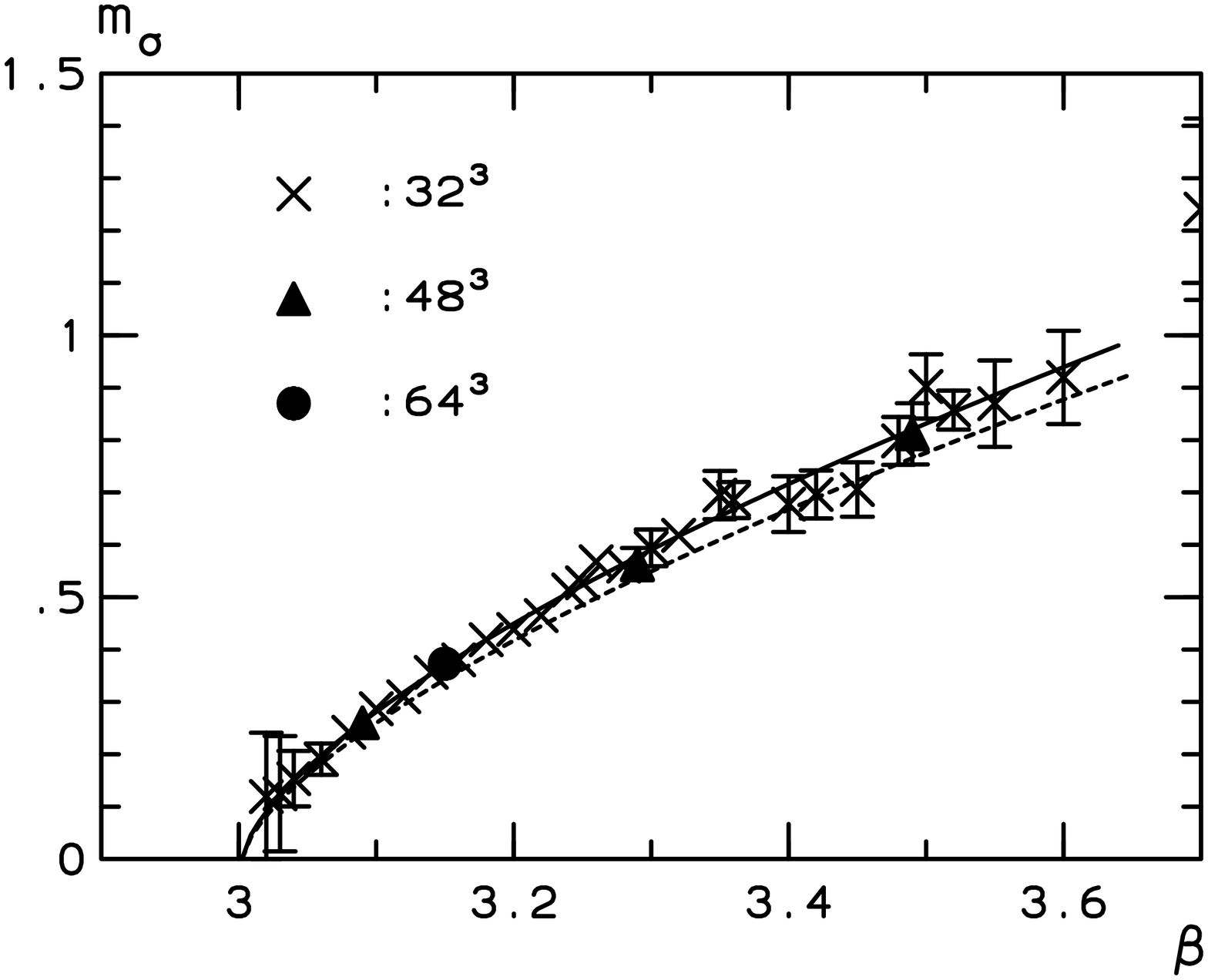,angle=270,width=6.7cm}\hfill}
\vspace{-1cm}
\label{fig:tens}
\caption[a]{Vortex tension in FZS (left) and the scalar
mass in the XY model (right), plotted against $\beta=1/\kappa_\xy$.
The continuous lines show power law fits, and, for comparison, the
dashed lines show the fits transferred from the other plot.}
\vspace{-2mm}
\end{figure*}

Our starting point is the lattice GL model in the
London limit:
\be
{\cal L}_{\rm GL}=
\frac{1}{2}\sum_{i<j}F_{\vec{x},ij}^2
+\kappa\sum_i s\left(\theta_{\vec{x}+i}-\theta_{\vec{x}} -qA_{\vec{x},i}
\right).
\ee
Here {$A_{x,i}$} is a real-valued gauge field, $\theta_x$ is the
spin angle variable, and 
$F_{\vec{x},ij}$ is the non-compact plaquette.
We use the Villain form hopping term
\be
s(x) = -{\rm ln} \sum_{k=-\infty}^{\infty} 
\exp\big[- \half (x-2\pi k)^2\big].
\ee
We shall study the following 2 limiting cases of the GL model:

i) Let $\kappa \rightarrow \infty$ and define $\beta = 4 \pi^2/q^2$.
Now the GL model becomes the {\em Frozen Superconductor}  (FZS):
\be
Z_{\fzs}(\beta)=\sum_{\{I_{\vec{x},i}\}} 
\exp\Big(- \frac{\beta}{2} \sum_{\vec{x},i>j} \Box_{\vec{x},ij}^2\Big).
\ee
Here $\Box_{\vec{x},ij}= I_{\vec{x},i}+I_{\vec{x}+i,j}-I_{\vec{x}+j,i}-
I_{\vec{x},j}$, and the link variables {$I_{\vec{x},i}$} take integer values.

ii) Let $q \rightarrow 0$, and the GL model becomes the
XY model with the Villain action
\be
Z_{\xy}(\kappa)=\int {D\theta} \exp\Big(-\kappa \sum_{\vec{x},i} 
s (\theta_{\vec{x}+i}-\theta_{\vec{x}})\Big).
\ee
The frozen superconductor and the XY-Villain model are {\em exactly dual}
to each other with the identification $\beta = 1/\kappa$, i.e.
\be
  Z_{\xy}(\kappa) = Z_{\fzs}(\beta={1}/{\kappa}).
\ee
This relation is valid at infinite volume; on a finite volume there
are corrections proportional to the surface area of the volume.  
For a proof of this relation see \cite{results}.

\section{CRITICAL OBSERVABLES}

The XY-Villain model has a symmetry breaking transition at $\kappa=\kappa_c
\approx 0.333068(7)$ \cite{results}.  Because of the exact duality, FZS must have
a transition at $\beta_c = 1/\kappa_c$, which is of XY model universality,
and the phases are related as follows:

\vspace{2mm}
\centerline{\begin{tabular}{ll}
\hline
XY model:  &  $\leftrightarrow$  FZS: \\
  symmetric $\kappa < \kappa_c$
\hspace*{-4mm}
& $\leftrightarrow$ 
  superconducting  $\beta > \beta_c$
\\
  broken   $\kappa > \kappa_c$
& $\leftrightarrow$ 
  Coulomb
 $\beta < \beta_c$
\\
\hline
\end{tabular}}

\subsubsection*{Vortex tension}

The duality implies that the XY model scalar
correlation function equals the FZS Abrikosov-Nielsen-Olesen
``vortex operator.''  In particular, the XY model scalar mass 
$m$ equals the tension ${\cal T}$ of the vortex line between
a monopole-antimonopole pair in FZS.
Thus, in the symmetric/superconducting phase of the XY/FZS model
we should find
\be
  m(\kappa) = {\cal T}(1/\kappa) \propto |\kappa -\kappa_c|^{\nu_\xy},
\ee
and in the broken/Coulomb phases $m = {\cal T} = 0$. 

This is indeed the case: in Fig.~\ref{fig:tens} we show the results
for ${\cal T}$ and $m$.  As predicted by duality, ${\cal T} = m$
within the statistical errors, and the tension critical exponent is
$0.672(9)$, compatible with $\nu_\xy$.

\subsubsection*{Photon mass}
The most natural observable for the FZS model (and the one usually
measured in high-$T_c$ superconductor experiments) is the photon mass
$m_\gamma = 1/\lambda$, the inverse of the penetration depth.
The duality relates $\lambda$ to the correlation length of the
Noether current operator in the XY-Villain model.
Parametrizing the critical behaviour of $m_\gamma$ with the exponent
$\nu'$, we have $m_\gamma \propto |\beta - \beta_c|^{\nu'} $ in the
superconducting phase.  The theoretical prediction is $\nu' = \nu_\xy$
\cite{herbut}.

\begin{figure}
\centerline{\psfig{file=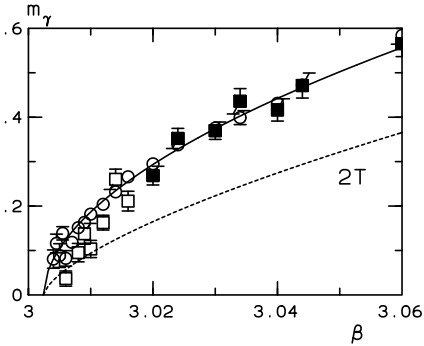,angle=270,width=7cm}}
\vspace{-1cm}
\label{fig:mg}
\caption[a]{Photon mass measured from the FZS model in close
proximity to the critical point.  The continuous line is a power law
fit, and the dashed line shows the curve $2\times {\cal T}$.}
\vspace{-5mm}
\end{figure}

Fig.~\ref{fig:mg} shows the $m_\gamma$ values measured from the
plaquette correlation functions in FZS.  A power law fit yields $\nu'
\approx 0.54$, {\em not compatible with the prediction from the
duality}.  However, this value agrees with (one of)
the experimental results \cite{experiment1} 
and Monte Carlo simulations of the GL theory \cite{results}.
The dashed line shows $2\times{\cal T}$.  Since
the photon operator we use couples to two vortices, and the measured
$m_\gamma > 2\times{\cal T}$, we conclude that the {\em observed
$m_\gamma$ shows pre-asymptotic behaviour}, and the true asymptotic
value is $m_\gamma = 2\times{\cal T}
\propto (\beta-\beta_c)^{\nu_\xy}$, which agrees with the duality.


\subsubsection*{Anomalous dimension}
At the critical point ($\beta=\beta_c$), the photon (plaquette) correlation
function is predicted to have a large anomalous dimension $\eta_A
= 1$ \cite{herbut}
\be
  \langle \Box(-\vec p)\Box(\vec p)\rangle
  \propto |\vec p\,|^{\eta_A}, \mbox{~~~when~~} 
	|\vec p| \rightarrow 0.
\ee
This is indeed what we observe: at $\beta_c$ the correlation
function in FZS shows a clear power law dependence, with the power
$\eta_A = 0.98(4)$.

\section{CONCLUSIONS}

Because of the exact duality between the frozen superconductor (FZS)
and the XY model, the critical observables in FZS must behave
according to the corresponding XY model critical exponents.  We have
measured several observables in the frozen superconductor, and we duly
find the behaviour predicted by the duality (for quantities other than
reported here, see \cite{results}).
The sole exception is the {\em photon
mass} $m_\gamma$, which appears to scale
with an exponent substantially smaller than the XY model prediction.
Exactly analogous behaviour has been observed superconductor
experiments and in GL theory simulations.  

Since the observed behaviour of $m_\gamma$ differs from predictions
even when the duality is exact, the similar discrepancies observed in
other studies do not mean that the duality hypothesis is not valid in
real superconductors.  The apparent incorrect scaling
behaviour may be due to the large anomalous dimension $\eta_A$ of the
photon propagator.

\end{document}